# EXPERIMENTS AND NUMERICAL RESULTS ON NONLINEAR VIBRATIONS OF AN IMPACTING HERTZIAN CONTACT. PART 1: HARMONIC EXCITATION

E. RIGAUD AND J. PERRET-LIAUDET

*Laboratoire de Tribologie et Dynamique des Systèmes UMR 5513,*

*36 avenue Guy de Collongue, 69134 Ecully cedex, France.*

Short headline : HARMONICALLY EXCITED HERTZIAN CONTACT

19 pages

17 figures

Final version

## Summary

The purpose of this paper is to investigate experimental and numerical dynamic responses of a preloaded vibro-impacting Hertzian contact under sinusoidal excitation. Dynamic response under random excitation is analysed in the second part of this paper. A test rig is built corresponding to a double sphere-plane contact preloaded by the weight of a moving cylinder. Typical response curves are obtained for several input levels. Time traces and spectral contents are explored. Both amplitude and phase of harmonics of the dynamic response are investigated.

Linearised resonance frequency and damping ratio are identified from the almost linear behaviour under very small input amplitude. Increasing the external input amplitude, the softening behaviour induced by Hertzian nonlinear stiffness is clearly demonstrated. Resonance peak is confined in a narrow frequency range. Jump discontinuities are identified for both amplitude and phase responses. Forced response spectrum exhibits several harmonics because of nonlinear Hertzian restoring force. Numerical simulations show a very good agreement with experimental results.

For higher input amplitude, system exhibits vibro-impacts. Loss of contact non-linearity clearly dominates the dynamic behaviour of the vibroimpacting contact and leads to a wide frequency range softening resonance. Spectral content of the response is dominated by both the first and the second harmonics. Evolution of the experimental downward jump frequency versus input amplitude allows the identification of the nonlinear damping law during intermittent contact. Simulations of the vibroimpacting Hertzian contact are performed using a shooting method and show a very good agreement with experimental results.

# 1. Introduction

Many mechanisms and mechanical devices use Hertzian contacts to transform motions and forces and to ensure rotation or translation motions. Under operating conditions, these contacts are often excited by dynamic normal forces superimposed on a mean static load. These excitation forces can be generated externally to the contact. They can also be generated by internal sources as roughness induced vibrations. Undesirable vibration responses can lead to excessive wear, contact fatigue and noise generation. Furthermore, in many mechanisms and mechanical devices such as gears and rolling element bearings, clearances introduced through manufacturing tolerances are necessary to ensure good functioning. Under excessive excitations, contacts can exhibit undesirable vibroimpact response leading to surface damage and excessive noise. In this context, study of the dynamic behaviour of fundamental Hertzian contacts including possible loss of contact is an essential stage in the understanding of the dynamics of more complex mechanical systems.

Considering vibrations of Hertzian contact excited by sinusoidal force, numerous papers present theoretical studies concerning the primary resonance which occurs without loss of contact [1, 2, 3]. Nayak modeled a preloaded Hertzian contact problem using the harmonic balance method [1]. The primary resonance exhibits a nonlinear softening behaviour induced by the nonlinear contact stiffness. Nevertheless, bending of the resonance peak remains in a narrow frequency range. Other theoretical works related to the primary resonance of Hertzian contact include that of Hess and Soom [2]. They analysed the reduction of the average friction coefficient induced by the dynamic response, using the multiple scales method. Perret-Liaudet and Sabot analysed the primary resonance including vibroimpact responses using a shooting method in relation with a continuation method [3]. The effect of the contact loss non-linearity is strong as the softening resonance is established in a much wider frequency range. Finally, theoretical description of the 2-subharmonic resonance and of the 2-superharmonic resonance is achieved by Perret-Liaudet [4, 5].

Partial experimental results concerning the primary resonance are presented by Carson and Johnson [6]. They used an original test rig consisting of two rolling contact discs, one of them being regularly corrugated. Sabot *et al.* [7] experimentally studied a ball normally preloaded by a moving rigid mass. They clearly exhibited the softening primary resonance when no loss of contact occurs and analysed mechanical sources of damping.

In this paper, experimental dynamic behaviour of a preloaded double sphere-plane Hertzian contact under sinusoidal excitation is investigated. Dynamic responses are investigated in detail including vibroimpact responses. Comparison with theoretical results permits us to conclude on the main characteristics of the system associated to both Hertzian non-linearity and contact loss non-linearity. After description of the studied dynamic model in section 2, we present the used test rig and the experimental procedure in section 3. Experimental and theoretical results are presented in section 4.

## 2. The studied dynamic model

### 2.1 Equation of motion

The single-degree-of-freedom impact oscillator under study is shown in Figure 1. A moving rigid mass $m$ is kept in contact with a flat surface and loaded by a static normal load $F_S$. Assuming Hertzian contact law, the nonlinear restoring contact force can be derived from material properties and contact geometry [8]. When the system is excited by a purely normal zero-mean force $F(t)$ superimposed on the static load, the equation of motion may be written as:

$$m\ddot{z}+c\dot{z}+k[H(z)z]^{3/2}=F_s+F(t) \qquad (1)$$

where $z$ is the normal displacement of the rigid mass $m$ measured such as $z<0$ corresponds to loss of contact, $c$ is a damping coefficient, $k$ is a constant given by the Hertzian theory and $H$ is the Heaviside step function. For convenience, the damping is assumed to be a constant at all times in equation (1) but other laws are introduced later.

When zero-mean excitation force is assumed, the static contact compression $z_s$ is given by the following equation:

$$z_s = \left(\frac{F_s}{k}\right)^{2/3} \tag{2}$$

Introducing the linearised contact natural frequency $\Omega$ and the damping ratio $\zeta$ given by:

$$\Omega^2 = \left(\frac{3k}{2m}\right) z_s^{1/2} \qquad \zeta = \frac{c}{2m\Omega} \tag{3, 4}$$

and rescaling equation (1) by letting:

$$q = \frac{3}{2}\left(\frac{z - z_s}{z_s}\right), \qquad \tau = \Omega t, \qquad f = \frac{F}{F_s} \tag{5, 6, 7}$$

the dimensionless equation of motion is obtained as follows:

$$\ddot{q} + 2\zeta\dot{q} + \left[H\left(1+\frac{2}{3}q\right)\left(1+\frac{2}{3}q\right)\right]^{3/2} = 1 + f(\tau) \tag{8}$$

In this equation, overdot indicates differentiation with respect to the dimensionless time $\tau$. It should be noted that loss of contact now corresponds to the inequality:

$$q < -3/2 \tag{9}$$

In this paper, a purely harmonic excitation is considered. Then, the dynamic external normal force is:

$$f(\tau) = \sigma \sin(\varpi\tau) \tag{10}$$

where $\sigma$ controls the level of the excitation and $\varpi$ is the dimensionless excitation circular frequency.

**2.2 Contact damping force**

To describe the contact damping force $f_d$, one may assume also several viscous damping laws which can be expressed, considering the original model case as:

$$f_d = 2\zeta\left(1+\frac{2}{3}q\right)^n H\left(1+\frac{2}{3}q\right)\dot{q} \tag{11}$$

One can introduce a linear damping force (n = 0), or a damping force proportional to the contact radius (n = 1/2), or a damping force proportional to the elastic deformation and to the contact area (n = 1), or a damping force proportional to the elastic restoring force (n = 3/2). For these laws, it should be noted that damping acts only when contact. More complex damping contact laws have been introduced in preceding studies [9, 10, 11] but have not been investigated in this study.

## 3- Test rig and experimental procedure

### 3.1 Test rig

Test rig is displayed in Figure 2. The experimental studied system corresponds to a double sphere-plane Hertzian contact. A 100C6 steel ball is compressed between the horizontal plane surfaces of two 100C6 steel thick discs which are rigidly fixed to a heavy rigid frame of a machine tool and a cylinder moving like a rigid body.

Ball diameter is 25.4 mm and its weight is 70 g. Double sphere plane Hertzian contact is loaded by the weight of the moving cylinder. Its mass is m=11.4 kg, corresponding to a static load $F_s$=mg=110 N. The moving cylinder is held by six titanium thin stems connected to the rigid frame in order to prevent lateral displacements and rotations. Then, only vertical motion of the cylinder is authorised. Furthermore, stems also allow regulation of the cylinder verticality.

Compliance of a rough and weakly loaded sphere-plane contact obtained experimentally may be different from the theoretical compliance supplied by the Hertz equation [12]. Planes were ground to obtain roughness Ra<0,4 µm in order to avoid this problem. Ball roughness is also weak (Ra<0.03 µm), so that, asperities are quite smaller than contact deflection and contact area. Finally, only dry contact is considered. Surfaces are cleaned before each run to remove grease from contact.

The experimental system can be modelled by a two degrees of freedom nonlinear dynamic system. However, the ball mass is negligible with regard to that of the moving cylinder, and it can easily shown that, for sufficiently low frequencies, this model is equivalent to the previously defined single degree of freedom system (1) [7]. Constant k of the preceding restoring elastic force expression is deduced from the double sphere-plane Hertzian contact. Assuming identical mechanical properties for the ball and the discs leads to:

$$k = \frac{E\sqrt{R}}{3\sqrt{2}(1-\nu^2)} \qquad (12)$$

where E is the Young modulus (210 Gpa), $\nu$ is the Poisson's ratio (0.29) and R is the ball radius (12.7 mm).

Then, theoretical characteristics of the experimental system are:

$$k = 5.98 \ 10^9 \ \text{N.m}^{-3/2} \qquad (13)$$

$$z_s = 7 \ \mu m \qquad (14)$$

$$f_0 = \frac{\Omega}{2\pi} = 232 \ \text{Hz} \qquad (15)$$

The second natural frequency of the double sphere-plane contact has been calculated and confirmed experimentally. Below this frequency (5500 Hz), the single degree of freedom system is justified.

Maximum contact pressure induced by static load has been calculated using Hertz theory ($p_0$=1.2 GPa) to ensure an elastic contact.

### 3.2 Instrumentation and acquisition

Contact is normally excited by a vibration exciter connected to the moving cylinder and suspended with four springs. Sinusoidal input is applied to the moving cylinder and superimposed on the static load $F_s$. For this end, a signal generator and a power amplifier are used.

A piezoelectric force transducer is mounted between the vibration exciter and the moving cylinder to measure the excitation force F(t). The vertical response $\ddot{z}(t)$ of the cylinder is measured by a piezoelectric accelerometer. Normal force N(t) and tangential forces T(t) transmitted to the frame through the contact are measured by a piezoelectric tri-axis force transducer. Classical charge amplifiers are used for all responses.

Regulation of the cylinder verticality leads to a quasi-perfectly normal load. During experimental measures, we check that tangential forces are negligible and remain always lower than 1 % of the normal force transmitted to the frame. Considering the experimental system and the transducer stiffness (8000 N/µm), the force measurement bandwidth is 0-7 kHz.

The input force and the dynamic responses are displayed on a four-channel storage oscilloscope. Spectral contents (amplitude and phase) are measured with a real time spectrum analyser using a sampling rate of 4096 samples over the frequency bandwidth. Then the frequency resolution is never unless than 0.25 Hz for all spectral quantities. Each harmonic of signal is analysed using a lock-in amplifier. This one is based on a phase sensitive detection to single out the components of the signal (frequency, amplitude, and phase).

**4- Forced dynamic response to a sinusoidal excitation**

All the experimental data obtained show a near perfect similarity between the fluctuating part of the normal force and the vertical acceleration of the moving cylinder. Measured correlation coefficients are always up to 99 %. Then, for convenience, only results associated to the normal force are presented.

**4.1 Dynamics without contact loss**

When external input amplitude is very small ($\sigma$ =0.03 %), contact dynamic behaviour is almost linear, even if the resonance curve is slightly asymmetrical (see Figure 3). No jump phenomenon occurs and harmonic normal force is observed. Linearised contact frequency measured from these experimental data ($f_0$=233.4 Hz) is close to the predicted natural frequency since the relative error is less than 0.5 % (see equation (15)). Analysis of the amplitude response curve of $H_1$ allows to estimate an equivalent viscous damping ratio close to $\zeta$=0.5 %. This result is coherent with preceding studies [7].

For higher external input amplitudes, experimental normal force N(t) presents not only $H_1$ component associated to the input frequency $\varpi$ but also $H_2$ component associated to the second harmonic $2\varpi$. The others harmonic components remain always negligible. Amplitude and phase of the two first harmonics ($H_1$ and $H_2$) are displayed in Figures 4 and 5, for two external input amplitudes. They are obtained increasing or decreasing the external input frequency, such as stationary process can be assumed at each frequency.

For an external input amplitude $\sigma$ =0.6 %, the system exhibits a nonlinear softening behaviour since the amplitude frequency response curve is bent to frequencies lower than the linearised contact natural frequency. Distortion of the phase response curve is also clearly demonstrated. The bending of the resonance curve leads to multi-valued amplitude and phase responses for $\varpi < 1$. Two stable solutions and one unstable solution exist leading to a pair of saddle node bifurcations and hence, to jump discontinuities. Jump discontinuities occur for both harmonic components $H_1$ and $H_2$. The downward jump frequency $\varpi_d$ corresponds to the phase resonance. Decreasing the external frequency, the phase angle of $H_1$ component relative to the excitation varies from 180° to 0° with a jump discontinuity from 90° to 0°. Phase angle of $H_2$ component is twice the phase angle of $H_1$ component. During data acquisition, tracking of these phase angles allows the prediction of the downward jump discontinuity before it occurs.

Normal response is very high for relatively small input amplitude, so that external input amplitude $\sigma$ =1 % leads to dynamic response just below loss of contact. Frequencies of upward jump and downward jump discontinuities have decreased. They are close to $\varpi_u$=0.984 and $\varpi_d$=0.957. The downward jump frequency agrees very well with the theoretical value predicted considering the backbone curve ($\varpi_d$=0.953) [1]. As we can see in Figures 3 and 4, $H_2$ component amplitude becomes non-negligible and reaches 17 % of $H_1$ component amplitude in the resonance peak. This behaviour results from the asymmetrical characteristic of the nonlinear restoring Hertzian force around the static equilibrium.

Minimum and maximum peak amplitudes of the normal force are displayed in Figure 6. Time history of the normal force at the resonance peak is displayed in Figure 7. It shows hardening behaviour in compression ($N_{max} = F_s + 1.3\ F_s$) and softening behaviour in extension ($N_{min} = F_s - 0.92\ F_s$). Spectrum of the normal force is displayed in Figure 8. $H_1$ component appears up to 1 although loss of contact does not occur. Consequently, resonance peak cannot be accurately predicted from analytical methods only taking into account the fundamental harmonic such as the harmonic balance method and the multiple scales method [1, 2]. $H_2$ component is close to 0.2 and the higher components are negligible.

Finally, the softening behaviour, well known through numerical results [1, 2, 3], is clearly demonstrated experimentally. Nonlinear behaviour associated to Hertzian contact is rather weak since the resonance curve is confined in a narrow frequency range close to the linearised contact frequency.

### 4.2 Dynamics with intermittent contact loss

For this set of experimental results, it is important to say that good repeatability was always observed.

Experimental results show that a small input amplitude is likely to induce intermittent loss of contact. When $\sigma$ >1.2 %, loss of contact occurs as the normal force reaches 100 % of the static load, that is to say the vertical acceleration of the moving cylinder reaches the gravity acceleration. Vibro-impact response leads to contact fatigue and undesirable noise (until 80 dB 1 meter far from the test rig). Figures 9 and 10 display the frequency response curves for increasing input amplitudes up to 1.2 % of the static load. Loss of contact non-linearity clearly dominates the dynamic behaviour. It strongly bends the frequency response curve to low frequency (softening behaviour). For instance, in the case of an input amplitude close to 7 % of the static load, upward jump frequency is equal to $\varpi_u$=0.909 and peak resonance occurs at $\varpi_d$=0.560 (nearly half the linearised contact frequency), so that, dynamic vibro-impact response can be established over a wide frequency range. Furthermore, the spectral content of the response is now much richer. $H_2$ component amplitude reaches up to 80 % of $H_1$ component amplitude in the peak resonance. We can also observe irregularities on the frequency response curve just before the downward jump occurs. These ones certainly result from the excitation of the second mode, inducing a ball motion between the upper and the lower planes.

Figure 11 displays the experimental downward jump frequency $\varpi_d$ versus the input amplitude $\sigma$. We can observe three behaviours. Just below the linearised contact frequency, (0.957< $\varpi_d$ <1) no loss of contact occurs and the downward jump frequency decreases slowly. When vibroimpacts occur, the downward jump frequency suddenly decreases over a relatively large frequency range and then slowly decreases (0.675< $\varpi_d$ <0.957). Then, the downward jump frequency suddenly decreases again ($\varpi_d$<0.675).

Figure 12 displays the normal force time histories for various input amplitudes. Input frequency is close to the downward jump frequency. Responses are asymmetrical and momentary loss of contact can be clearly observed. Then, for $\sigma$ =1.4 % (and $\varpi$ =0.9), loss of contact occurs during 40 % of the overall time, for $\sigma$ =3 % ($\varpi$ =0.76), loss of contact occurs during 60 % of the overall time, and for $\sigma$ =7 % ($\varpi$ =0.55), loss of contact occurs during 75 % of the overall time. For the last case, cylinder fly period reaches 5.8 ms. Furthermore, instantaneous transmitted normal force reaches 600 % of the static load for the presented time histories. However, plastic deformation has not been observed on surfaces.

Finally, we can observe that the last time history of the normal force exhibits the influence of the second mode characterised by the ball motion between the upper and the lower planes.

**4.3 Theoretical results**

We have used a classical numerical time integration explicit scheme (central difference) for achieving dynamic time histories of theoretical responses. A specific computing method devoted to nonlinear problems is used to obtain the primary resonance, that is to say a shooting method with continuation technique (see for example [13, 14]).

In the case of an external input amplitude just below loss of contact, theoretical prediction of the primary resonance is displayed in Figure 13 and compared to experimental results. As we can see, theoretical results are in very good agreement with experimental ones for both amplitude and phase response curves. The narrow frequency band of the softening resonance is confirmed. Damping ratio introduced in the numerical model (0.45%) is close to the one experimentally measured at very low excitation (see section 4.1). By introducing different damping laws (see equation (11) ), we have numerically observed that they don't affect the primary resonance curves. This result is coherent with preceding study [7].

Figure 14 displays the primary resonance exhibiting the strong nonlinear behaviour induced by loss of contact. The wide frequency range softening resonance observed is in good agreement with experimental results. However, the introduced damping ratio is higher than the one experimentally measured at very low excitation. In the contrary of the preceding case, the damping contact law strongly influences the downward jump frequency prediction.

Downward jump frequencies $\varpi_d$ versus input amplitude $\sigma$ (see equation (11) ) are displayed in Figure 15 for different damping laws. Results are compared with experimental ones displayed in Figure 11. On the contrary to third experimental behaviour observed for $\varpi_d<0.675$, the two first experimental behaviours can be explained by theoretical results. As expected, for a constant input amplitude, downward jump frequency associated to damping proportional to elastic restoring force is higher than the one associated to linear damping during contact. A linear damping law during contact is not suitable in contrast with nonlinear model. Even if the predicted and the experimental behaviours are similar, predicted softening resonance peaks are wider than experimental ones, for all the introduced damping laws during contact. Several explanations may be done as follows:

- Other sources of damping may exist becoming more and more significant with the amplitude response.

- Other kinds of damping law during contact may be more adapted to treat the theoretical model.

- Some interactions between the first and the second modes may affect the dynamic behaviour of the system resulting in occurrence of other bifurcations.

Anyway it appears necessary to carry on efforts of research in order to obtain precise knowledge of damping mechanisms during impacts. Actually, in our opinion, damping in vibroimpact conditions is not adequately modelled at the present time.

Finally, time traces of the steady state normal force response are given in Figures 16 and 17 illustrating respectively a case without loss of contact and a case with. Theoretical results are in a good agreement with experimental ones (see Figures 7 and 12) despite adjustment of damping ratio in vibroimpact conditions. In particular, when impacts occur, flight time ratio is correctly predicted, compared to in contact time and maximum amplitude. Nevertheless, increasing the damping ratio is necessary to adjust levels, even if a damping proportional to elastic restoring force is introduced.

## 5- Conclusion

An original test rig has been built and improved in order to analyse vibrations of a double sphere-plane preloaded Hertzian dry contact excited by a purely harmonic input normal force. Theoretical and experimental results concern the primary resonance. Experimental linearised contact frequency and theoretical value are very close (233 Hz). Damping ratio measured with small input amplitude is very low (0.5 %).

For input amplitude lower than 1 % of the static load, no loss of contact occurs. Experimental dynamics clearly demonstrates the softening behaviour of Hertzian contacts well known through numerical results.

The resonance curve is bent to frequencies lower than the linearised contact natural frequency, leading to saddle-node bifurcations and jump discontinuities. We have observed that the dynamic response is dominated by both the $H_1$ component associated to the input fundamental frequency $\varpi$ and the $H_2$ component associated to the second harmonic $2\varpi$. Jump discontinuities happening for both components $H_1$ and $H_2$ and the associated phases are clearly shown. But we can conclude that the Hertzian non-linearity remains rather weak since the resonance peak is confined in a narrow frequency range close to the linearised contact frequency. This result is confirmed by theoretical analysis.

As the damping is low, a small input amplitude is likely to induce vibroimpacts (around 1.2% of the mean static load). Loss of contact nonlinearity clearly dominates the dynamic behaviour as it strongly bends the frequency response curve to low frequencies (softening behaviour). Actually, resonance peak is established over a wide frequency range. Amplitude of the $H_2$ component becomes higher and higher and cylinder fly duration becomes longer and longer as input amplitude increases.

Specific computing methods devoted to nonlinear problems, i.e. shooting and continuation methods, have been used to treat the loss of contact non-linearity and to investigate theoretical dynamic response of the Hertzian contact. Theoretical results qualitatively agree very well with experimental ones. Quantitatively, slight discrepancies appear. Particularly, it is necessary to increase the theoretical damping ratio with the amplitude response even if a viscous damping proportional to the restoring elastic force is introduced. This result reveal a lack of comprehension of damping physical sources. So, in our opinion, some efforts of research are necessary to obtain precise knowledge of damping mechanisms in vibroimpact conditions.

**6- References**

## 7- Nomenclature

m rigid mass

c damping coefficient

k Hertzian constant

$F_s$ static load

F(t) excitation normal force

z(t) normal displacement

$z_s$ static contact compression

Ra surface roughness

R ball radius

E Young modulus

ν Poisson's ratio

Ω linearised natural circular frequency

$f_0$ linearised natural frequency

$p_0$ static contact pressure

ζ damping ratio

τ dimensionless time

q(τ) dimensionless normal displacement

f(τ) dimensionless excitation normal force

$f_d(\tau)$ dimensionless contact damping force

$\varpi$ dimensionless excitation frequency

$\sigma$ level of sinusoidal force

$N(t)$ transmitted normal force

$T(t)$ transmitted tangential force

$N_{max}$ maximum transmitted normal force

$N_{min}$ minimum transmitted normal force

$H_1$ first hamonic of the transmitted normal force

$H_2$ second harmonic of the transmitted normal force

$\varpi_u$ upward jump frequency

$\varpi_d$ downward jump frequency

Figure 1. Studied single degree of freedom oscillator.

Figure 2. Test rig. (1) Vibration exciter, (2) force transducer, (3) moving cylinder, (4) accelerometer, (5) ball, (6) tri-axis force transducer, (7) rigid frame.

Figure 3. Normal force vs. external frequency for $\sigma = 0.03$ %.

Figure 4. First harmonic amplitude and phase responses vs. external frequency for $\sigma = 0.6$ % (■) and $\sigma = 1$ % (○).

Figure 5. Second harmonic amplitude and phase responses vs. external frequency for $\sigma = 0.6$ % (■) and $\sigma = 1$ % (○).

Figure 6. Maximum and minimum peak amplitudes of the normal force vs. external frequency for $\sigma = 1$ %.

Figure 7. Time history of the normal force for $\sigma = 1$ % and $\varpi = 0.957$.

Figure 8. Amplitude spectrum of the normal force for $\sigma = 1$ % and $\varpi = 0.957$.

Figure 9. First harmonic amplitude vs. external frequency for $\sigma = 1.4$ %, $\sigma = 3$ % and $\sigma = 7$ %.

Figure 10. Second harmonic amplitude vs. external frequency for $\sigma = 1.4$ %, $\sigma = 3$ % and $\sigma = 7$ %.

Figure 11. Downward jump frequency vs. input amplitude.

Figure 12. Time histories of the normal force at the resonance peak for $\sigma = 1.4$ % ($\varpi=0.9$), $\sigma = 3$ % ($\varpi=0.76$), and $\sigma = 7$ % ($\varpi=0.57$).

Figure 13. Comparison between predicted ($\sigma = 0.6$ % and $\zeta = 0.45$ %) and experimental amplitude and phase responses without loss of contact. Thick line: predicted stable response; dotted line: predicted unstable response; upward experimental response (●) and downward experimental response (○).

Figure 14. Comparison between predicted ($\sigma = 0.6$ %, $\zeta = 0.45$ %, n=3/2 in equation (11) )and experimental responses with loss of contact. Thick line: predicted stable response; dotted line: predicted unstable response; thin line: experimental response.

Figure 15. Downward jump frequency vs. input amplitude for several damping laws (n=0, n=1/2, n=1, n=3/2 in equation 11)

Figure 16. Numerical time history of the normal force without vibroimpacts $\sigma = 1.1\,\%$, $\zeta = 0.45\,\%$ and $\varpi = 0.957$.

Figure 17. Numerical time history of the normal force with vibroimpacts for $\sigma = 7.8\,\%$, $\zeta = 1\,\%$ and $\varpi = 0.57$.

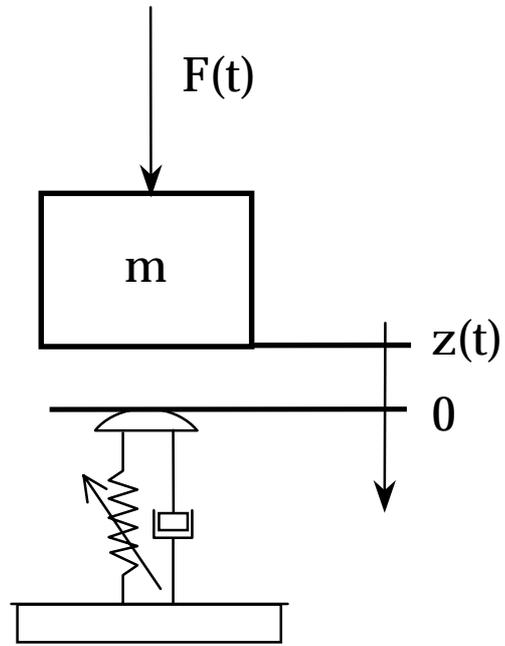

Figure 1. Studied single degree of freedom oscillator.

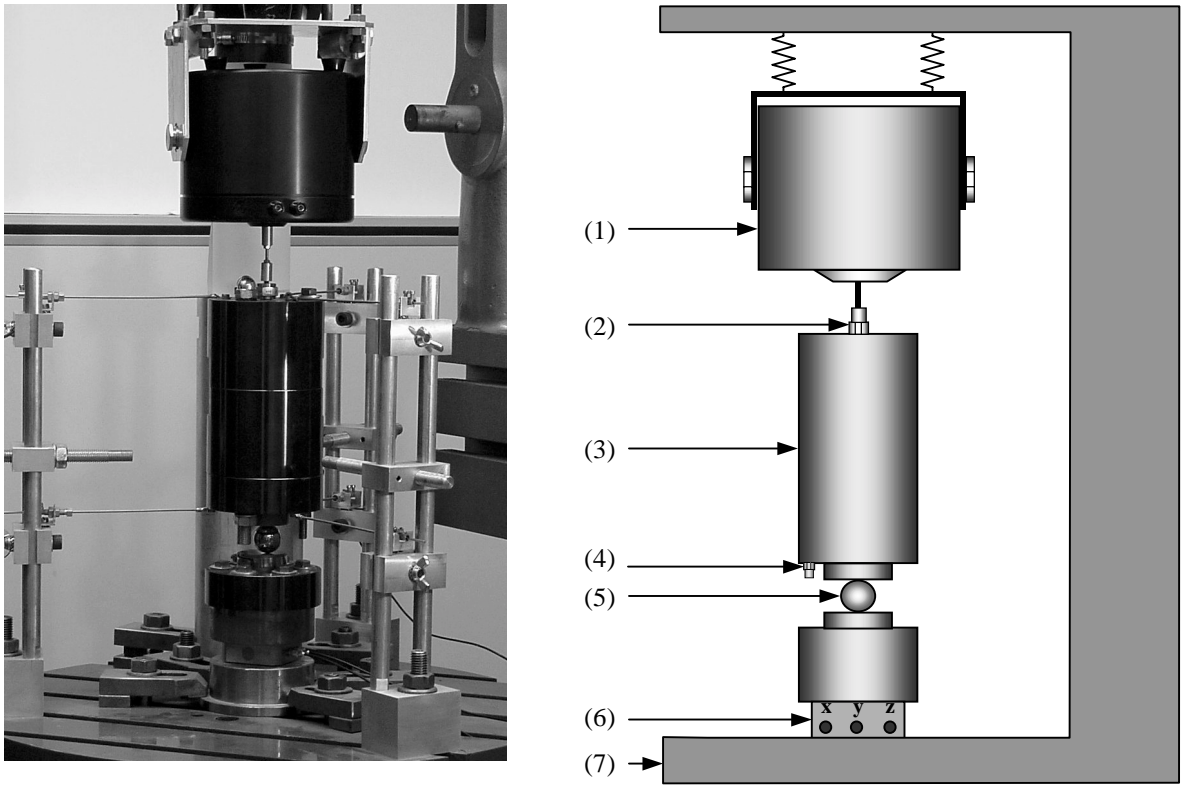

Figure 2. Test rig.

(1) Vibration exciter, (2) force transducer, (3) moving cylinder, (4) accelerometer, (5) ball, (6) tri-axis force transducer, (7) rigid frame.

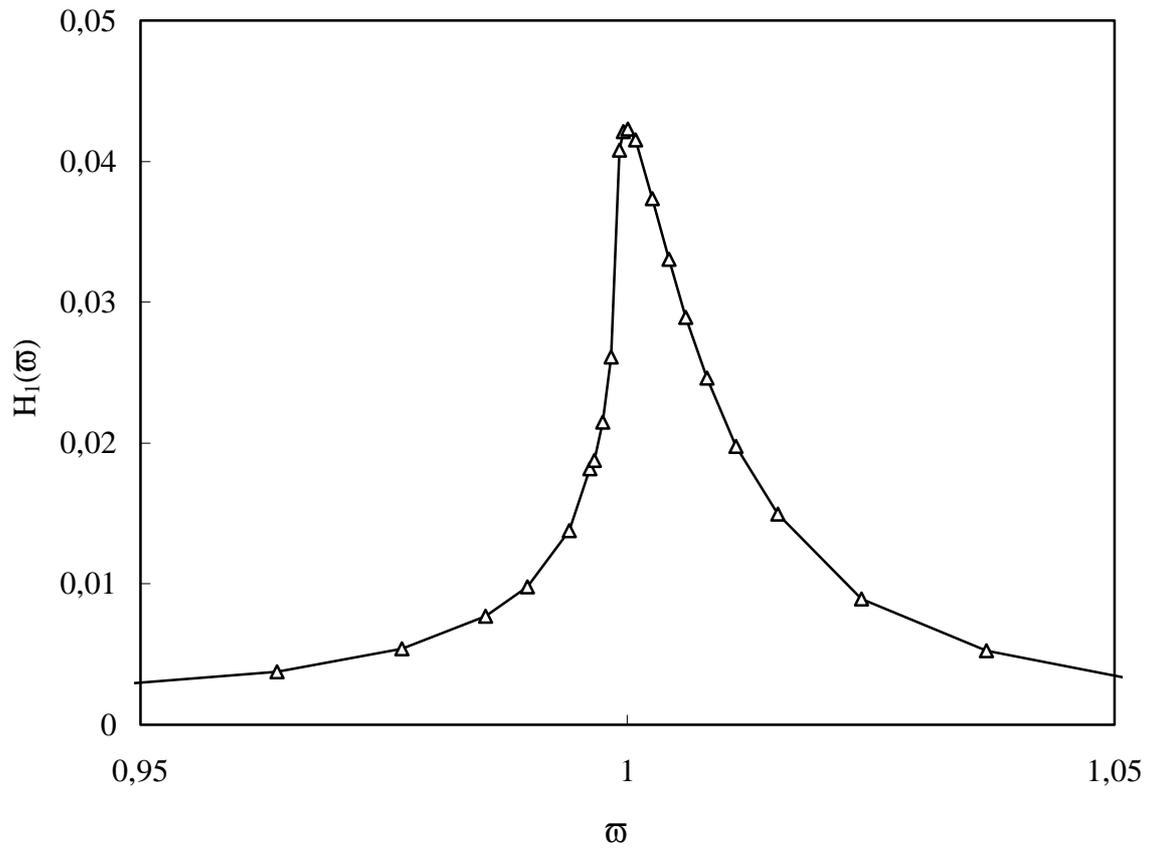

Figure 3. Normal force vs. external frequency for σ = 0.03 %.

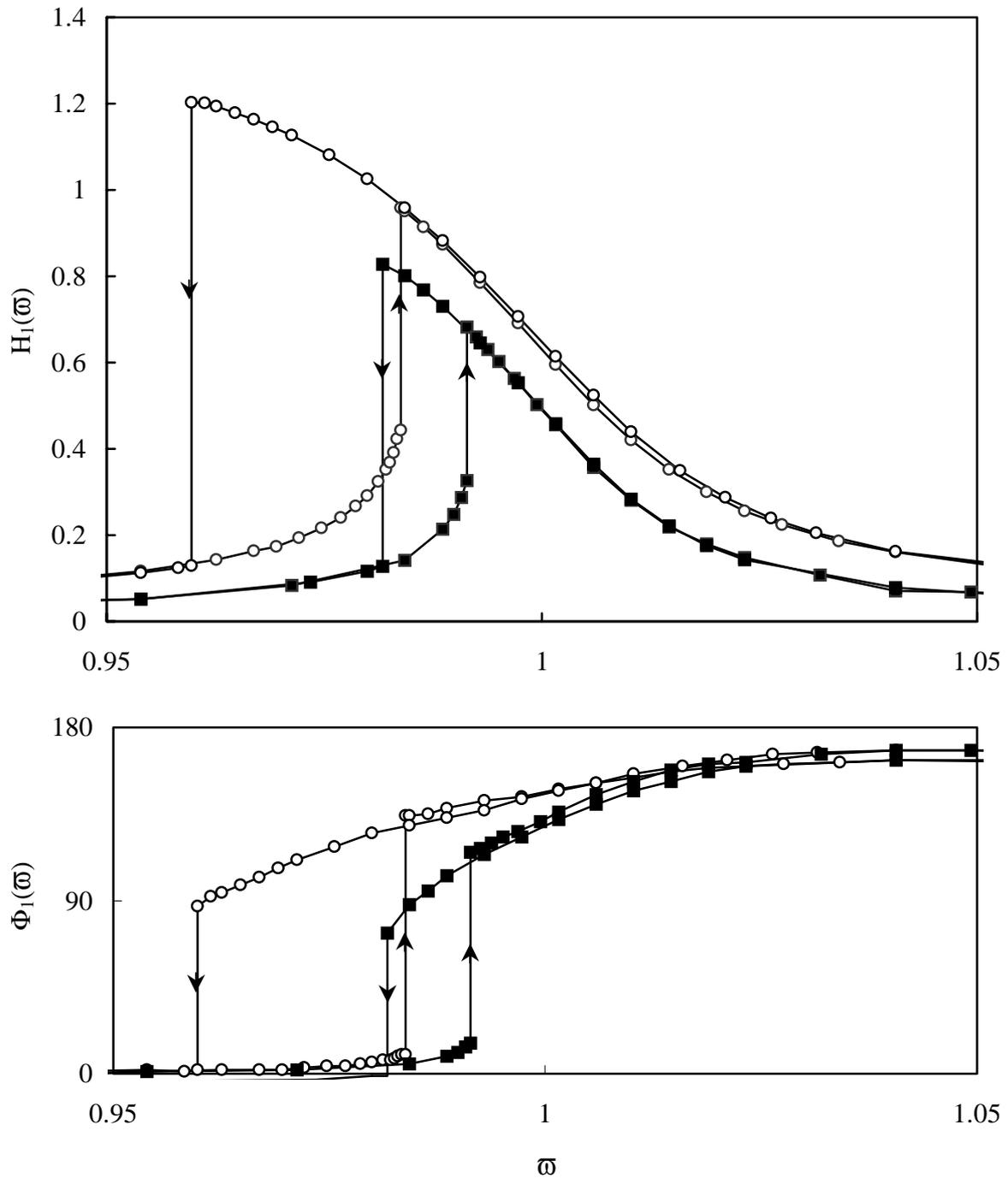

Figure 4. First harmonic amplitude and phase responses vs. external frequency for

σ = 0.6 % (■) and σ = 1 % (○).

E. RIGAUD AND J. PERRET-LIAUDET

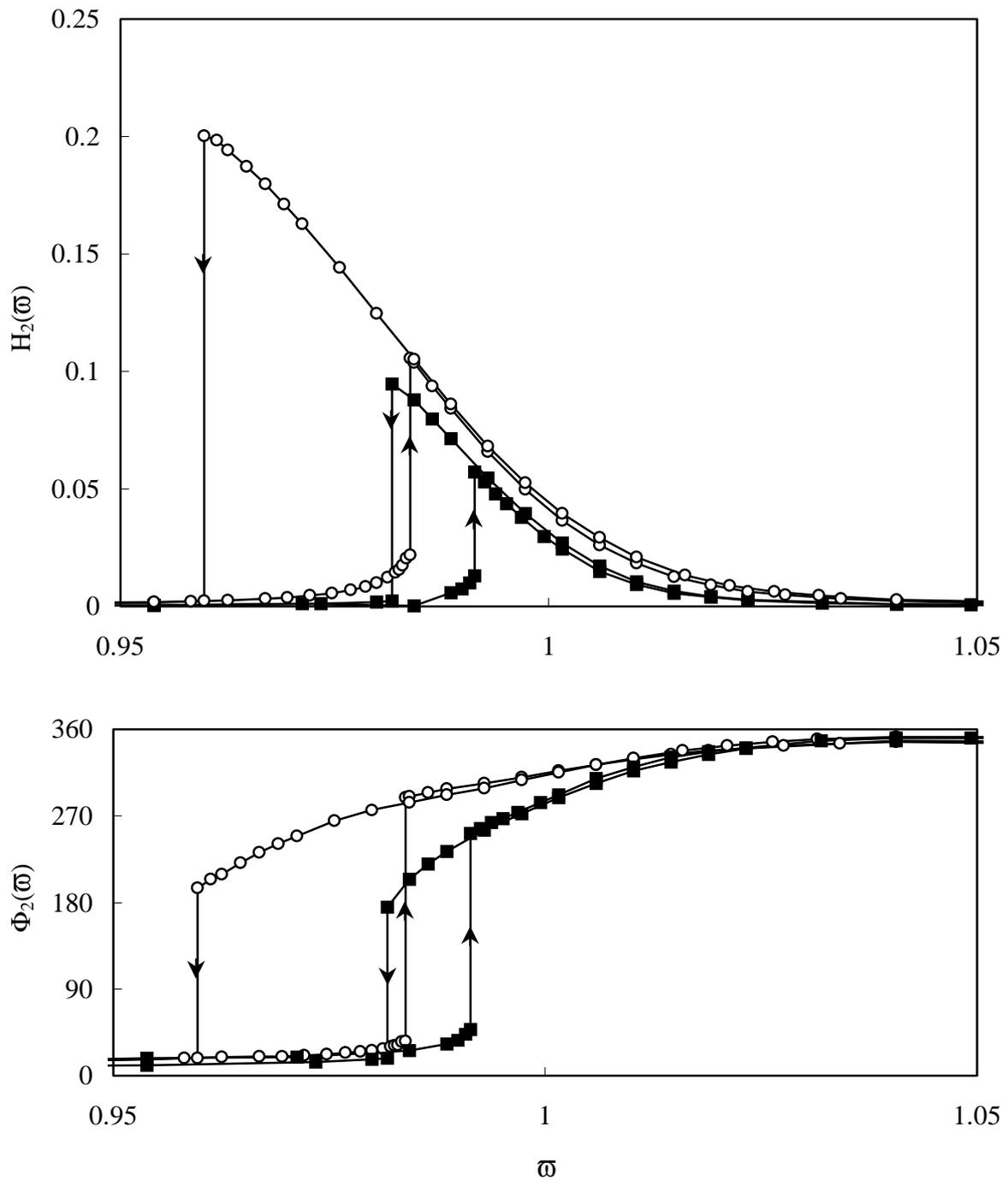

Figure 5. Second harmonic amplitude and phase responses vs. external frequency for

$\sigma = 0.6\ \%$ (■) and $\sigma = 1\ \%$ (○).

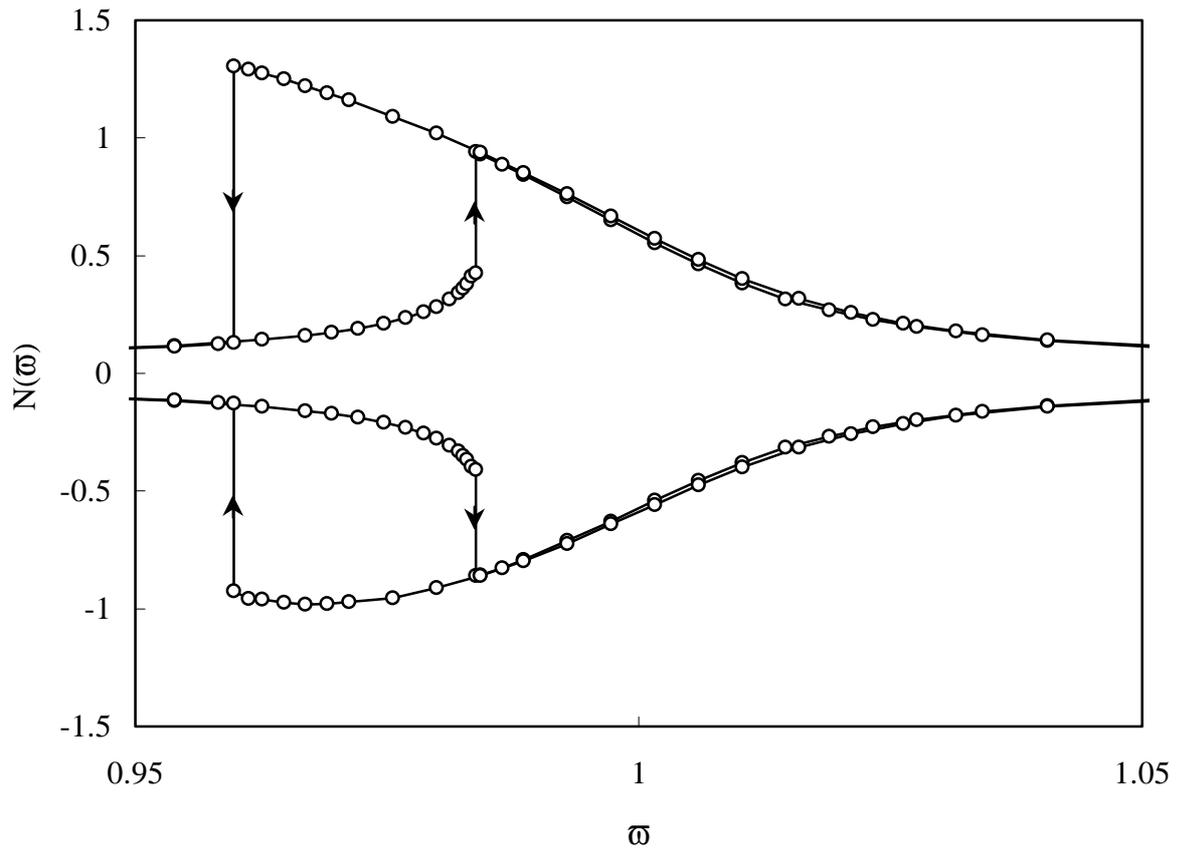

Figure 6. Maximum and minimum peak amplitudes of the normal force vs. external frequency

for $\sigma = 1\ \%$.

E. Rigaud and J. Perret-Liaudet

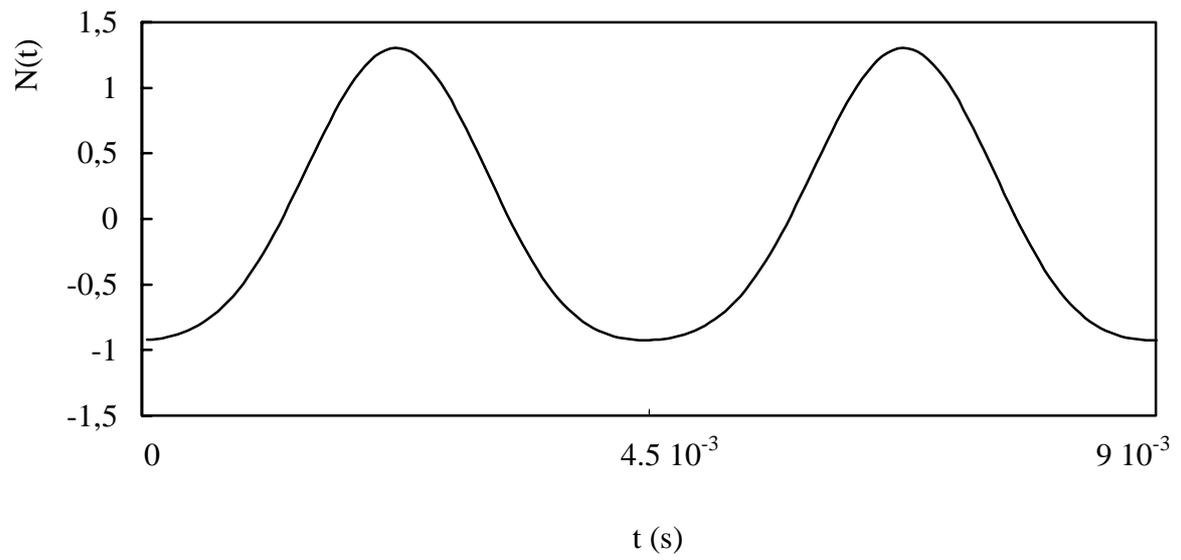

Figure 7. Time history of the normal force for σ = 1 % and $\varpi$ = 0.957.

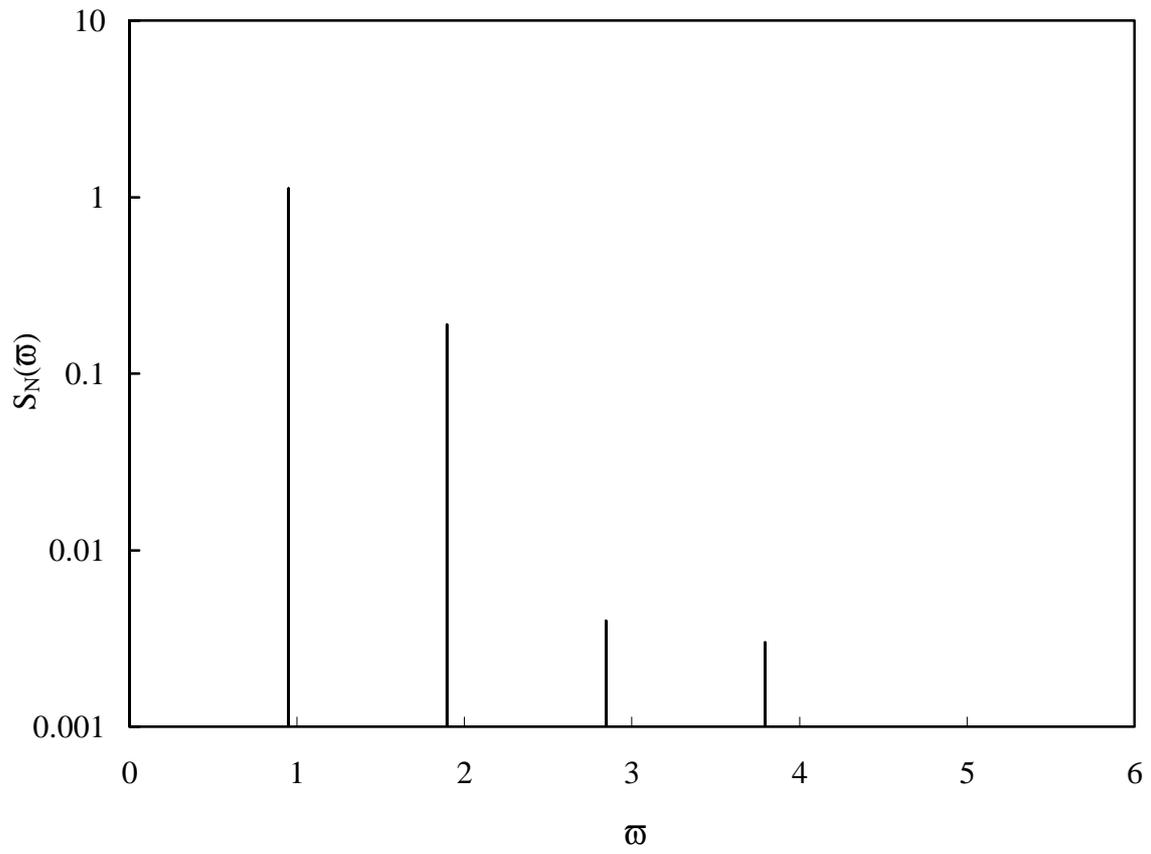

Figure 8. Amplitude spectrum of the normal force for σ = 1 % and $\varpi$ = 0.957.

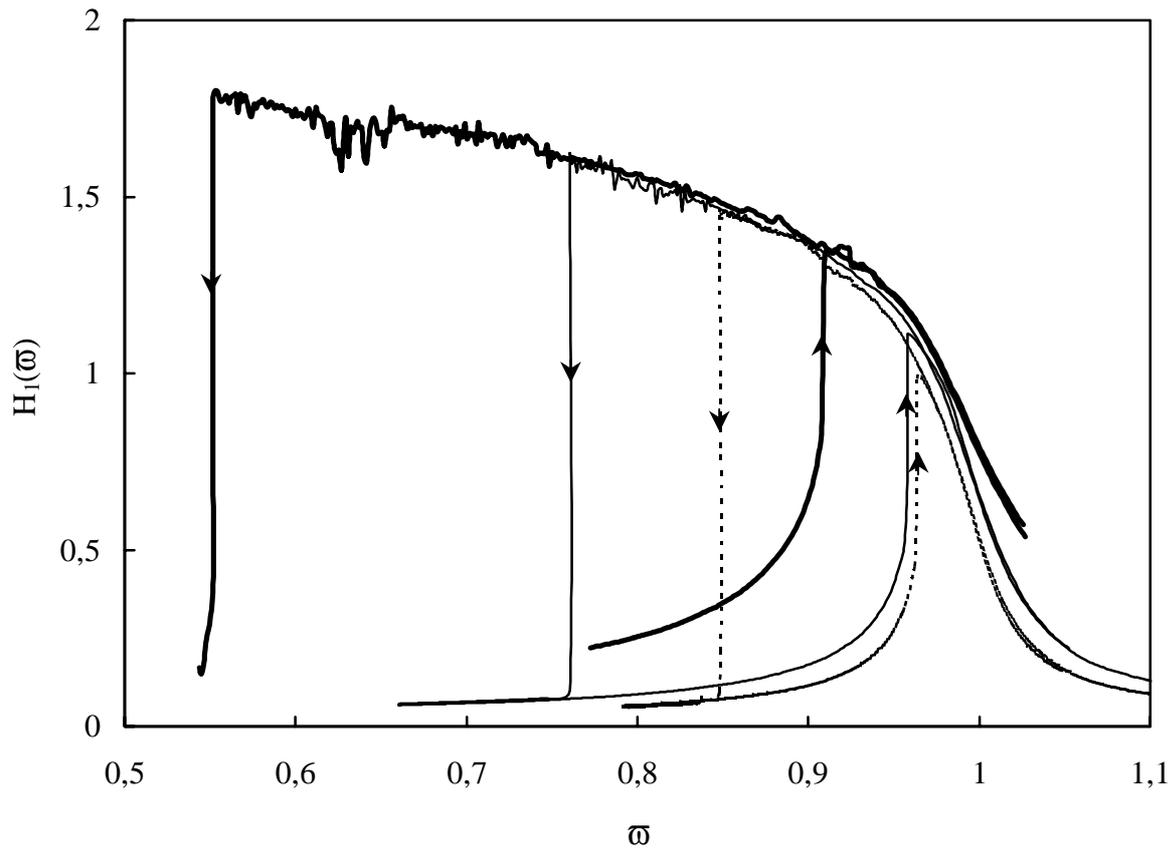

Figure 9. First harmonic amplitude vs. external frequency for σ =1.4 %, σ = 3 % and σ = 7 %.

E. RIGAUD AND J. PERRET-LIAUDET

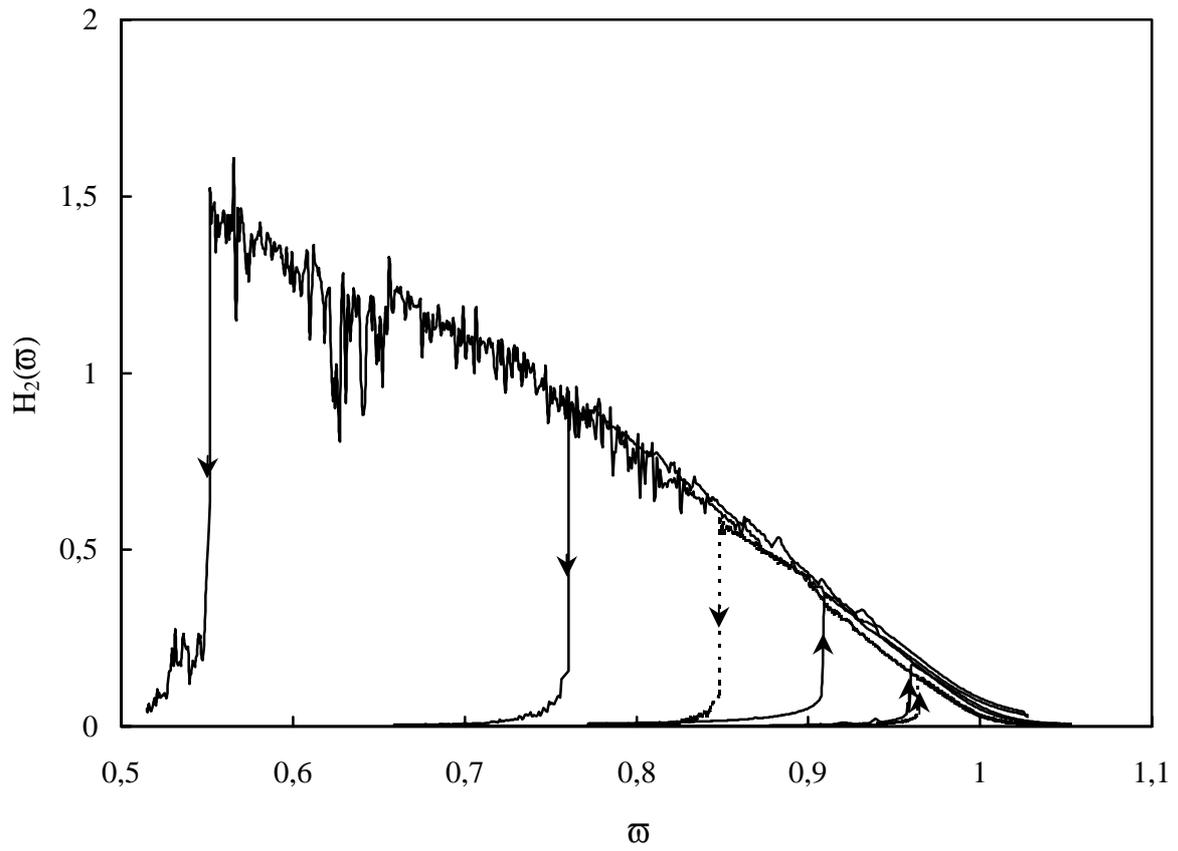

Figure 10. Second harmonic amplitude vs. external frequency for $\sigma = 1.4\ \%$, $\sigma = 3\ \%$ and $\sigma = 7\ \%$.

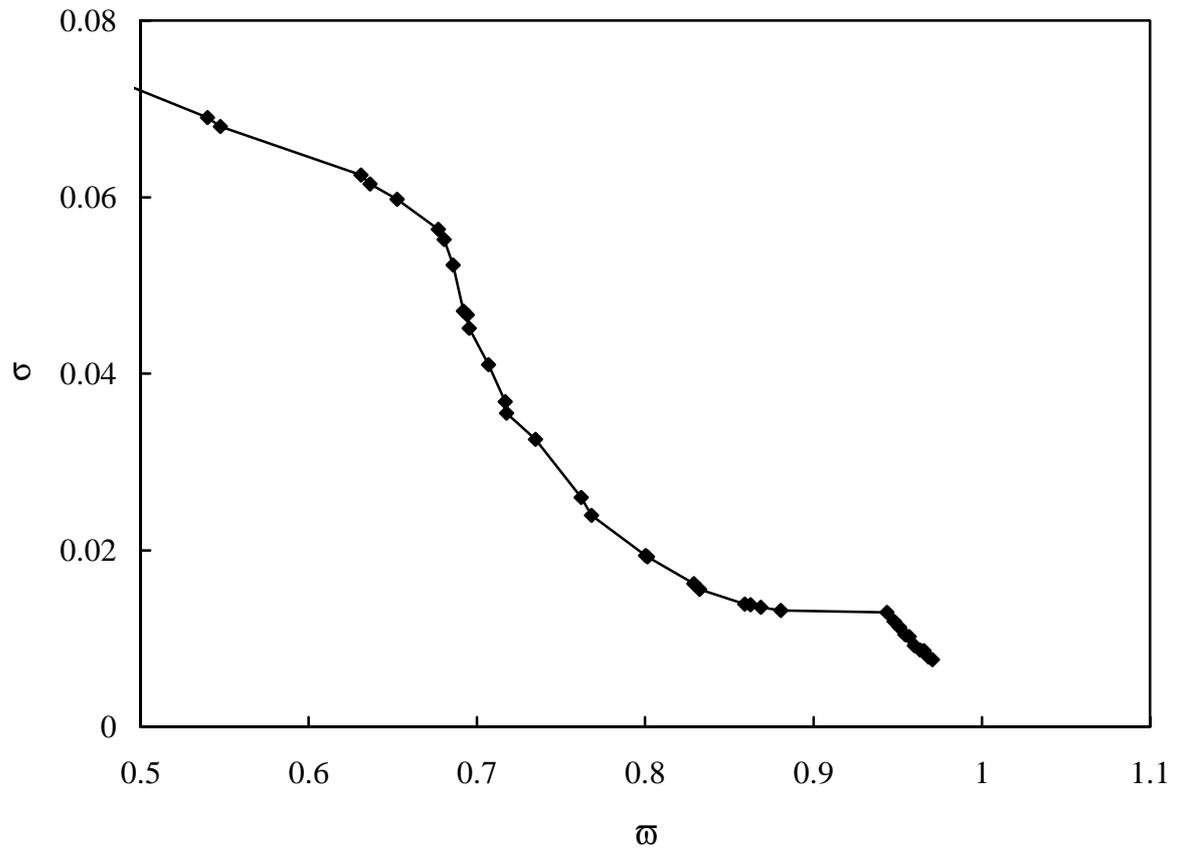

Figure 11. Downward jump frequency vs. input amplitude.

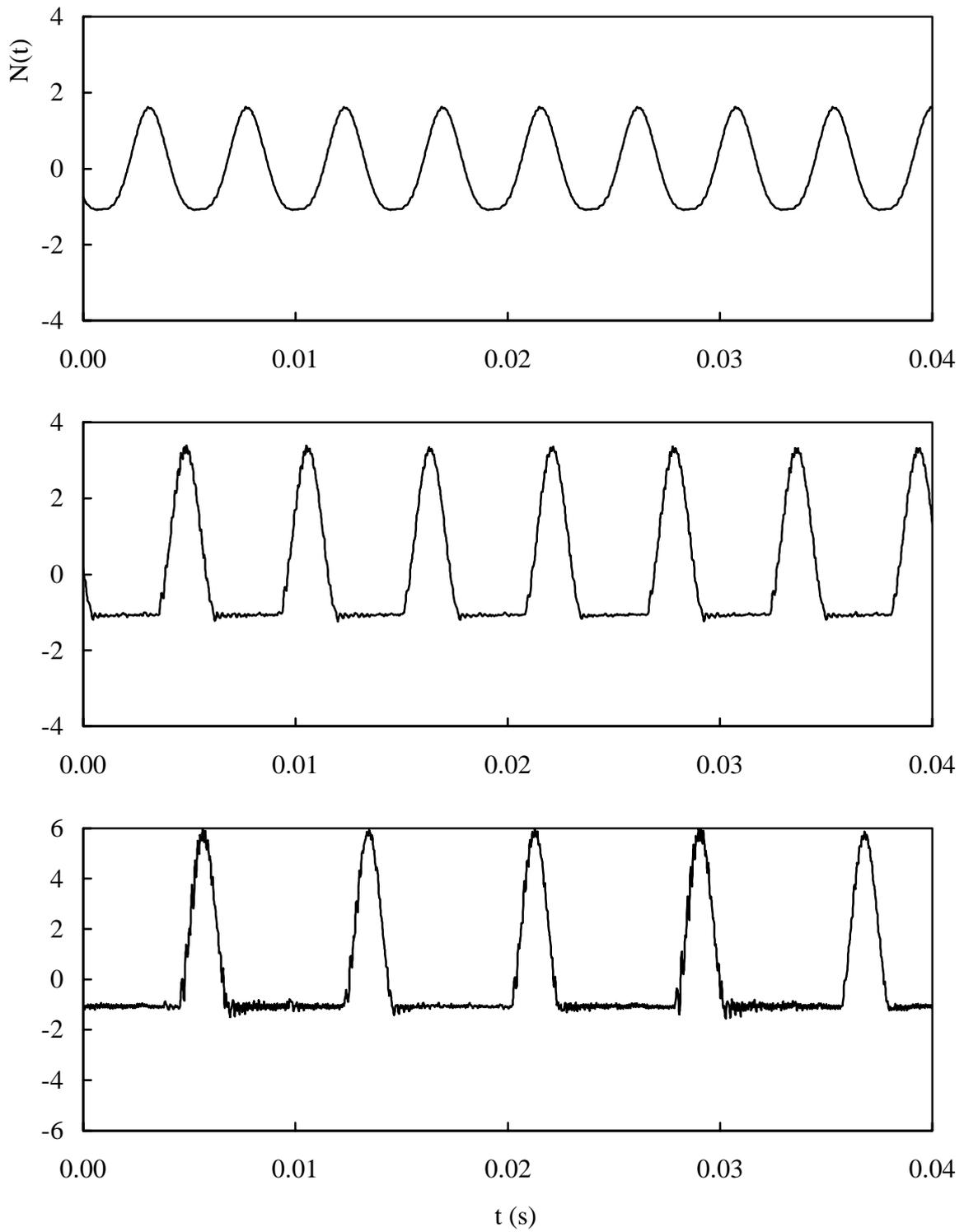

Figure 12. Time histories of the normal force at the resonance peak for σ = 1.4 % ($\varpi$=0.9), σ =3 % ($\varpi$=0.76), and σ = 7 % ($\varpi$=0.57).

E. RIGAUD AND J. PERRET-LIAUDET

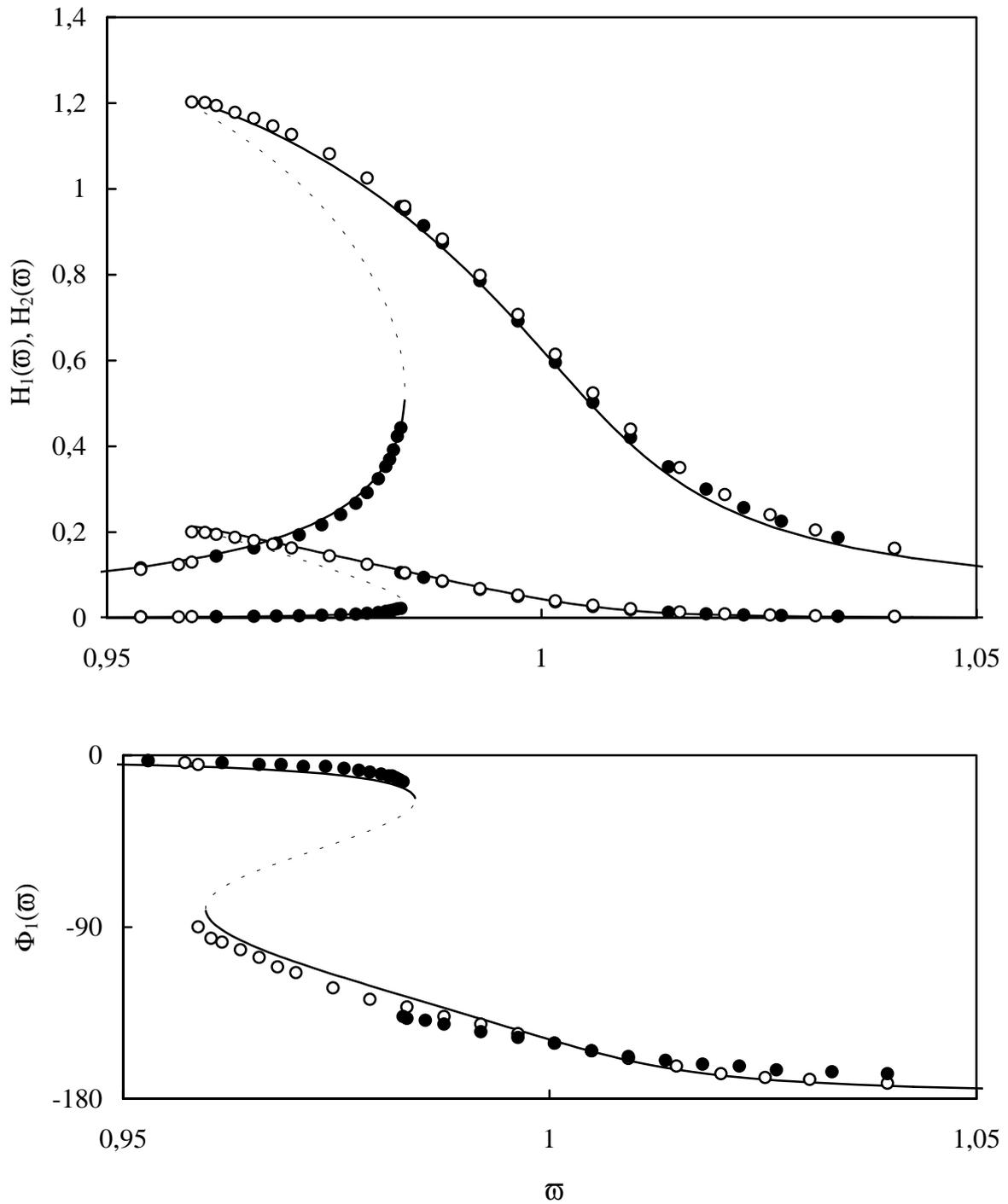

Figure 13. Comparison between predicted ($\sigma = 0.6$ % and $\zeta = 0.45$ %) and experimental amplitude and phase responses without loss of contact. Thick line: predicted stable response; dotted line: predicted unstable response; upward experimental response (●) and downward experimental response (○).

E. RIGAUD AND J. PERRET-LIAUDET

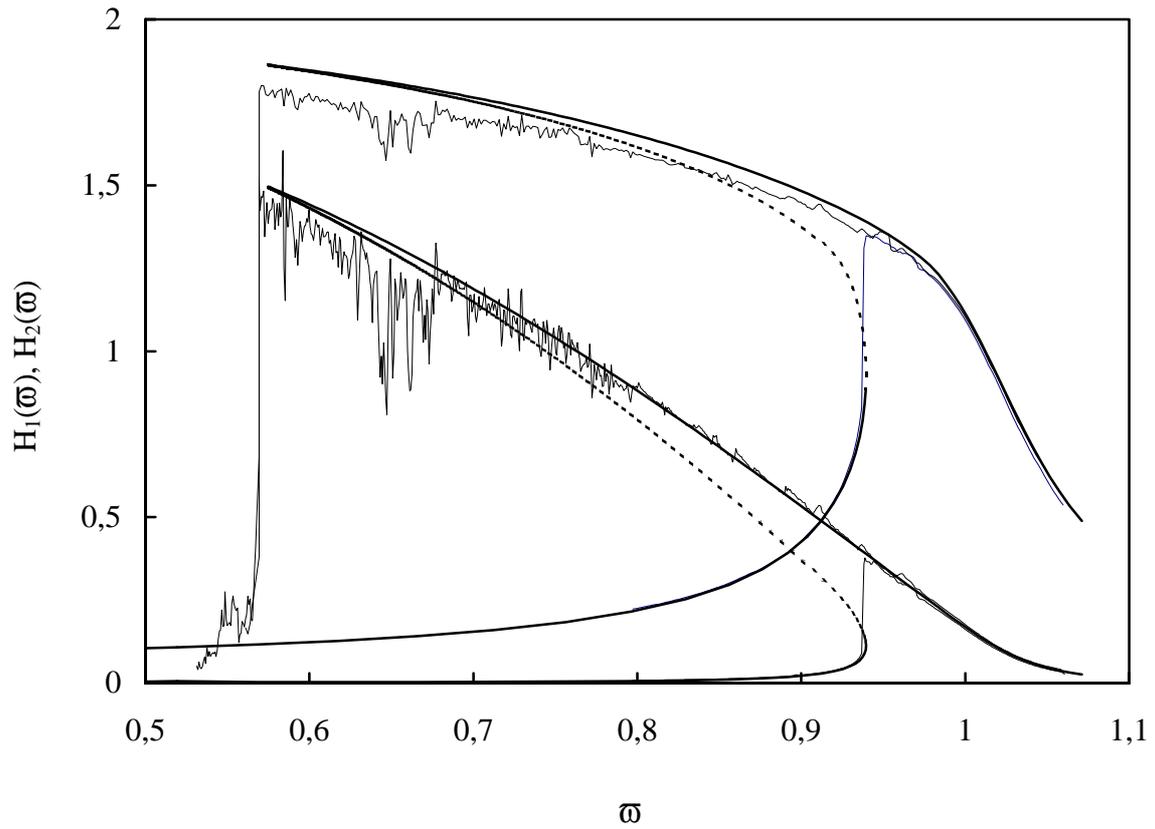

Figure 14. Comparison between predicted ($\sigma = 0.6$ %, $\zeta = 0.45$ %, n=3/2 in equation (11) )and experimental responses with loss of contact. Thick line: predicted stable response; dotted line: predicted unstable response; thin line: experimental response.

E. R<small>IGAUD</small> AND J. P<small>ERRET</small>-L<small>IAUDET</small>

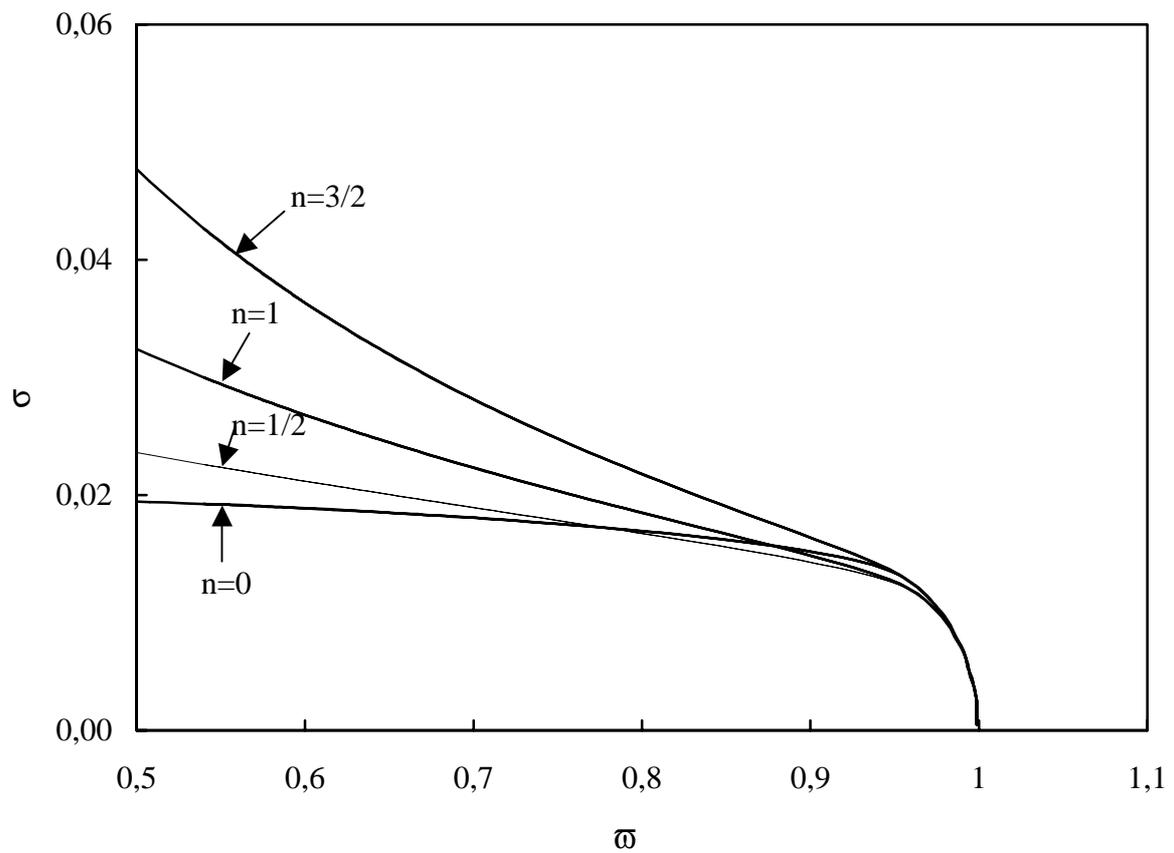

Figure 15. Downward jump frequency vs. input amplitude for several damping laws (n=0, n=1/2, n=1, n=3/2 in equation 11)

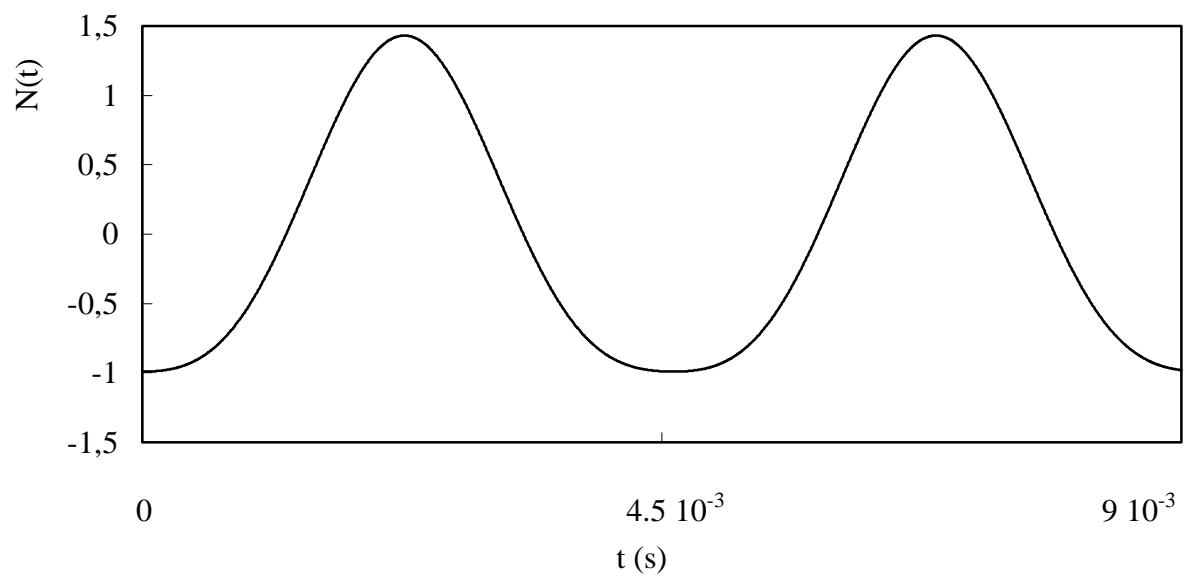

Figure 16. Numerical time history of the normal force without vibroimpacts $\sigma = 1.1$ %, $\zeta = 0.45$ % and $\varpi = 0.957$.

E. RIGAUD AND J. PERRET-LIAUDET

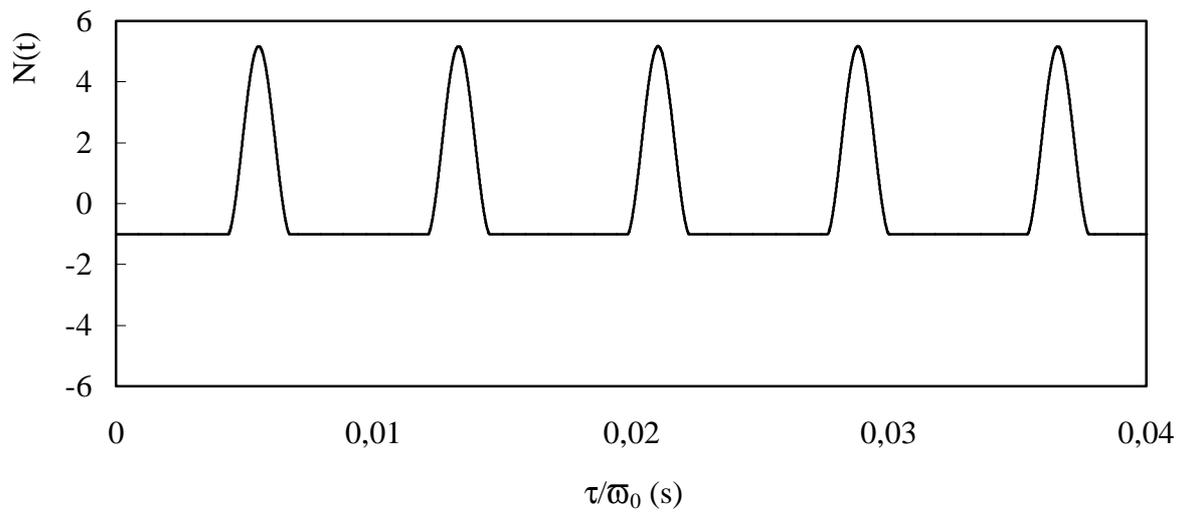

Figure 17. Numerical time history of the normal force with vibroimpacts for $\sigma = 7.8\%$, $\zeta = 1\%$ and $\varpi = 0.57$.

E. RIGAUD AND J. PERRET-LIAUDET